\documentclass{elsart}
\usepackage{amsmath, amssymb, amsxtra}
\usepackage{graphicx}
\journal{solid state communications}
\begin{document}
%
%
\begin{frontmatter}
\title{Kondo effect in side coupled double quantum-dot molecule}
%
%
\author[UA]{Gustavo A. Lara \corauthref{cor}}
\ead{glara@uantof.cl}
\corauth[cor]{Corresponding author.}
\author[UCN]{Pedro A. Orellana}
\ead{orellana@ucn.cl}
%
\author[UA]{Julio M. Y\'{a}\~{n}ez}
\ead{jyanez@uantof.cl}
\author[PUCR]{Enrique V. Anda}
\ead{anda@fis.puc-rio.br}
\address[UA]{Departamento de F\'{\i}sica,
            Universidad de Antofagasta,
            Casilla 170, Antofagasta, Chile}
\address[UCN]{Departamento de F\'{\i}sica,
             Universidad Cat\'{o}lica del Norte,
             Casilla 1280, Antofagasta, Chile}
\address[PUCR]{Departamento de F\'{\i }sica,
              P. U. Cat\'{o}lica do Rio de Janeiro,
              C.P. 38071-970, Rio de Janeiro,
              RJ, Brazil}
%
%
\begin{abstract}
  Electron tunneling through a double quantum dot molecule side
  attached to a quantum wire, in the Kondo regime, is studied. The
  mean-field finite-$U$ slave-boson formalism is used to obtain the
  solution of the problem. We found conductance cancelations  when
  the molecular energies of the side attached double quantum-dot
  cross the Fermi energy. We investigate the many body molecular
  Kondo states as a function of the parameters of the system.
\end{abstract}
%
%
\end{frontmatter}
%
%
\section{Introduction}

Quantum dots (QDs) are man-made nanostructures in which electrons
are confined in all three space dimensions \cite{JHW98}. Energy
and charge quantization results from this confinement. As both
features are present in real atomic systems, an useful analogy has
been used between ``real'' and ``artificial'' atomic systems.
Like-wise, a system of coupled QDs is called an ``artificial
molecule''. Enforcing this analogy, in QDs configurations Kondo
effect and Fano resonance are also present.

The Kondo effect is the name given to describe the resistivity
minimum for decreasing temperature in certain alloys with a minute
concentration of a magnetic impurity \cite{K69,K64}. It is
important to emphasize that in this case, the so-called
traditional Kondo effect, magnetic impurities act as scattering
centers, increasing the sample resistivity (for a review see
Ref.~\cite{H97}). The opposite behavior is found in the
Kondo effect in QDs, the so-called anti Kondo effect.
Experimentally \cite{GSMAMK98,GGKSMM98,COK98} and theoretically
\cite{GR88,NL88}, the situation considered usually consists of a
QD connected to two leads. In these configuration electrons
transmitted from one electrode to the other necessarily pass
through the QD.

The Kondo effect in QDs has been extensively
studied in the last years \cite{GSMAMK98,GGKSMM98,COK98}
(for a review see Ref.~\cite{PG04}).  The
experimental evidence has confirmed that many of the phenomena
that characterize strongly correlated metals and insulators, as it
is the case of the Kondo effect, are present in QDs. The QDs
allow to study systematically the quantum-coherence many-body
Kondo state, due to the possibility of continuous tuning the
relevant parameters governing the properties of this state, in
equilibrium and nonequilibrium situations. Recently Kondo effect
has been studied in double quantum dot molecule in series
\cite{GM99,AE01,JCM01,SG02,LOA03}. This system allows the study
of the many body molecular Kondo states in equilibrium and
nonequilibrium situation. The type of coupling between the QDs
determines the character of the electronic states and the transport
properties of the artificial molecule. In the tunneling regime, the
electronics states are extended across the entire system and form
a coherent state based on the bonding or anti-bonding levels of
the QDs.

An alternative configuration consists of a side-coupled QDs
attached to a perfect quantum wire (QW). This structure is
reminiscent of T-shaped quantum wave guides known as electron stub
tuners \cite{DRVRPM00} (see also Refs.~\cite{TIK02,GSG03}).
In this case, the QDs acts as scattering
centers in close analogy with the traditional Kondo effect
\cite{KCKS01,THCP02,ODGL03,FFA03,S03}.

Recent electron transport experiments showed that Kondo resonance
occurs simultaneously with the Fano resonance
\cite{GGHK00,ZGGKKSMM01}. Multiple scattering of traveling
electronic waves on
a localized magnetic state are crucial for a formation of both
resonances. The condition for the Fano resonance \cite{F61} to
appear is a presence of at least two scattering channels: the
discrete level and the broad continuum band \cite{TB93,NS94}.

In this work we study the transport properties of a double quantum
dot molecule side attached to a quantum wire in the Kondo regimen.
We use the finite-$U$ slave boson mean-field approach which was
initially developed by Koliar and Ruckenstein \cite{KR86} and used
later by Bing Dong and X. L. Lei to study the transport through
coupled double quantum dots connected in series to leads
\cite{DL01,DL01b,DL02,DL02b}. We found that the antiresonances of
the linear conductance reflect the spectral properties of the
artificial molecule. We investigate the many body molecular Kondo
states as a function of the parameters of the system.

\begin{figure}[h]
\centering
  \begin{picture}(220,100)(0,0)
    \thicklines
    \put(40,70){\line(1,0){140}}
    \put(110,70){\line(0,-1){60}}
    \put(110,10){\circle*{7}}
    \put(110,30){\circle*{7}}
    \put(100,30){\makebox(0,0){$\varepsilon_{d1}$}}
    \put(100,10){\makebox(0,0){$\varepsilon_{d2}$}}
    \put(120,50){\makebox(0,0){$t_{0}$}}
    \put(120,20){\makebox(0,0){$t_{c}$}}
    \put(50,75){\makebox(0,0)[b]{}}
    \put(70,75){\makebox(0,0)[b]{}}
    \put(90,75){\makebox(0,0)[b]{-1}}
    \put(110,75){\makebox(0,0)[b]{0}}
    \put(130,75){\makebox(0,0)[b]{1}}
    \put(150,75){\makebox(0,0)[b]{}}
    \put(170,75){\makebox(0,0)[b]{}}
    \put(110,90){\makebox(0,0)[b]{QW}}
    \put(150,20){\makebox(0,0)[l]{DQD}}
    \put(80,0){\dashbox{2}(60,40)}
    \put(10,70){\makebox(0,0)[r]{$\mu_{L}$}}
    \put(210,70){\makebox(0,0)[l]{$\mu_{R}$}}
    \put(40,70){\makebox(0,0)[r]{$.\, .\, .\, .\, .$}}
    \put(180,70){\makebox(0,0)[l]{$.\, .\, .\, .\, .$}}
    \multiput(50,70)(20,0){7}{\circle*{3}}
    \multiput(60,65)(20,0){6}{\makebox(0,0)[t]{}}
    \qbezier(0,50)(40,70)(0,90)
    \qbezier(220,50)(180,70)(220,90)
  \end{picture}
  \caption{Scheme of double quantum dot (DQD) attached
           to a lead (perfect quantum wire (QW)). The QW is
           coupled to the left ($L$) and right ($R$)
           ``reflectionless'' contacts.}
  \label{f:pendiente}
\end{figure}
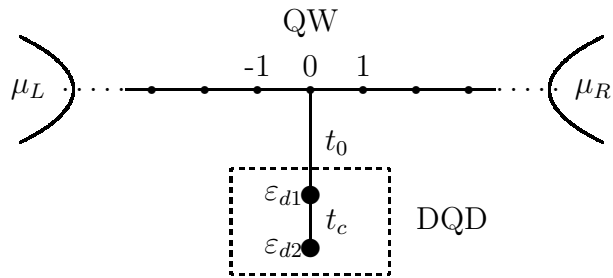

%
%
\section{Model}

Let us consider a double quantum dot (DQD) side coupled to a
perfect quantum wire (QW) (see Fig.~\ref{f:pendiente}). We adopt
the two-fold Anderson Hamiltonian, that is diagonalized using the
finite-$U$ mean field slave boson formalism. Each dot has a single
level $E_{i}$ (with $i = 1, 2$) and intradot Coulomb repulsion
$U$, and is coupled to each other with interdot tunneling coupling
$t_{c}$. The QW has an energy $\varepsilon_{i,\sigma}$ for site
and has a hopping parameter $t$.

In analogy with the infinite-$U$ slave boson approach
\cite{B76,B77,C84} the Hilbert space is enlarged at each site,
to contain in addition to the original fermions a set of four bosons
\cite{KR86} represented by the creation (annihilation) operators
$e_{i}^{\dag}$ ($e_{i}$), $p_{i, \sigma}^{\dag}$ ($p_{i, \sigma}$),
and $d_{i}^{\dag}$ ($d_{i}$) for the $i$-th dot, which act,
respectively, as projectors onto empty, single occupied (with spin
up and down) and doubly occupied electron states. Then the
Hamiltonian is written as,
\begin{equation}
  \begin{split}
     H =& \sum_{i,\sigma} \varepsilon_{i,\sigma} c_{i,\sigma}^{\dag}
     c_{i,\sigma} -t\sum_{i,\sigma} \left( c_{i,\sigma}^{\dag}
     c_{i+1,\sigma} + \text{H.c.} \right) -t_{0} \left( c_{0,
     \sigma}^{\dag} \tilde{Z}_{1,\sigma} f_{1,\sigma} + \text{H.c.}
     \right)                           \\
     & + \sum_{i=1}^{2} \sum_{\sigma} \left( \varepsilon_{di} + \lambda_{i,
     \sigma}^{(2)} \right) f_{i,\sigma}^{\dag} f_{i, \sigma}
     - t_{c} \sum_{\sigma} \left( f_{1, \sigma}^{\dag}
     \tilde{Z}_{1,\sigma}^{\dag} \tilde{Z}_{2,\sigma} f_{2,\sigma}
     + \text{H.c.} \right)             \\
     & + \sum_{i=1}^{2} \Big\{ U d_{i}^{\dag} d_{i} + \lambda_{i}^{
    (1)} \left( p_{i,\uparrow}^{\dag} p_{i,\uparrow} + p_{i,
    \downarrow}^{\dag} p_{i, \downarrow} + e_{i}^{\dag} e_{i} +
    d_{i}^{\dag} d_{i} - 1 \right)      \\
    & - \sum_{\sigma} \lambda_{i, \sigma}^{(2)}
    \left( p_{i, \sigma}^{\dag} p_{i,\sigma} +
    d_{i}^{\dag} d_{i} \right) \Big\} \, ,
  \end{split}
\end{equation}
where $c_{i,\sigma}^{\dag}$ ($c_{i,\sigma}$) is the creation
(annihilation) operator of a electron with spin $\sigma $ in the
$i$-th site on quantum wire; $f_{i,\sigma}^{\dag}$
($f_{i,\sigma}$) is the creation (annihilation) operator of a
electron with spin $\sigma $ in the i-th QD. The operator
$\tilde{Z}_{i, \sigma}$ is chosen to reproduce the correct $U \to
0$ limit in the mean field approximation,
\begin{equation}
  \tilde{Z}_{i, \sigma} = \left(1 - d_{i}^{\dag}d_{i} - p_{i,
  \sigma}^{\dag} p_{i, \sigma}\right)^{-1/2} \Bigl( e_{i}^{\dag}
  p_{i, \sigma} + p_{i, -\sigma}^{\dag} d_{i} \Bigr) \left(1
  - e_{i}^{\dag}e_{i} - p_{i, -\sigma}^{\dag} p_{i, -\sigma}
  \right)^{-1/2} \, .
\end{equation}
The constraint, i.e., the completeness relation $\sum_{\sigma}
p_{i, \sigma}^{\dag} p_{i, \sigma} + b_{i}^{\dag} b_{i} +
d_{i}^{\dag} d_{i} = 1$ and the condition for the correspondence
between fermions and bosons $f_{i,\sigma}^{\dag} f_{i,\sigma} =
p_{i,\sigma}^{\dag} p_{i,\sigma} + d_{i}^{\dag} d_{i}$,
have been incorporated with Lagrange multipliers
$\lambda_{i}^{(1)}$ and $\lambda_{i,\sigma}^{(2)}$ into the
Hamiltonian. Then we use the mean-field approximation in which all
the boson operators are replaced by their expectation value. These
expectation values and the Lagrange multipliers are then
determined by minimization of the free energy with respect to
these variables. So, the effective Hamiltonian is $H_{\text{eff}}
= H_{TB} + H_{B}$, where:
\begin{align}
  H_{TB} &= -t \sum_{i,\sigma} \left( c_{i,\sigma}^{\dag} c_{i+1,
  \sigma} + \text{H.c.} \right) - \sum_{\sigma} \tilde{t}_{0,\sigma}
  \left( c_{0, \sigma}^{\dag} f_{1,\sigma} + \text{H.c.} \right)
  \notag    \\
  & \quad + \sum_{i=1}^{2} \sum_{\sigma} \tilde{\varepsilon}_{di,
  \sigma} f_{i,\sigma}^{\dag} f_{i, \sigma} - \sum_{\sigma}
  \tilde{t}_{c, \sigma} \left( f_{1, \sigma}^{\dag} f_{2,\sigma}
  + \text{H.c.} \right)   \\
  H_{B} &= \sum_{i=1}^{2} \left\{ U d_{i}^{2} + \lambda_{i}^{(1)}
  \left( p_{i,\uparrow}^{2} + p_{i,\downarrow}^{2} + e_{i}^{2} +
  d_{i}^{2}  - 1 \right) - \sum_{\sigma}  \lambda_{i,\sigma}^{(2)}
  \left( p_{i, \sigma}^{2} + d_{i}^{2} \right) \right\} \, ,
\end{align}
with $\tilde{\varepsilon}_{di,\sigma} =  \varepsilon_{di} +
\lambda_{i, \sigma}^{2}$, $\tilde{t}_{0,\sigma} = t_{0} \langle
\tilde{Z}_{1, \sigma} \rangle $, and $\tilde{t}_{c,\sigma} = t_{c}
\langle \tilde{Z}_{1,\sigma} \tilde{Z}_{2,\sigma} \rangle$, and we
set $\varepsilon_{i,\sigma} = 0$ in the QW.
The Hamiltonian tight binding, $H_{TB}$, corresponds to an
effective  free-particle Hamiltonian on a lattice with spacing set
as unity, and whose eigenfunctions are expressed as Bloch
solutions
\begin{equation}
  \left| k \right\rangle = \sum_{j=-\infty}^{\infty}
  \text{e}^{\text{i}k\, j} \left| j \right\rangle \, ,
\end{equation}
where $\left| k \right\rangle$ is the momentum eigenstate and
$\left| j \right\rangle$ is a Wannier state localized at site $j$.
The dispersion relation associated with these Bloch states reads
$\omega = -2t \cos k $ .
Consequently, the Hamiltonian supports an energy band from $-2t$
to $+2t$ and the first Brillouin zone expands the interval
$[-\pi, \pi]$. The stationary states of the entire Hamiltonian
$H$ can be written as
\begin{equation}
  \left| \psi _{k} \right\rangle = \sum_{j=-\infty }^{\infty}
  a_{j}^{k} \left| j \right\rangle
  + \sum_{l=1}^{2} b_{l}^{k} \left| l \right\rangle \, ,
\end{equation}
where the probability amplitude to find the electron in the state
$k$ in the $j$-th site of the QW is the coefficient $a_{j}^{k}$,
and to find it in $l$-th QD is the coefficient $b_{l}^{k}$.

From $H_{TB} \left| \psi _{k} \right\rangle = \omega \left| \psi
_{k} \right\rangle $, the amplitudes $a_{j}^{k}$ and $b_{l}^{k}$
obey the following linear difference equations
\begin{subequations}
  \begin{align}
    \omega a_{j}^{k} &= -t(a_{j+1}^{k} + a_{j-1}^{k}) \, , \quad
    j \neq 0  \, , \label{dif1}                     \\
    \omega a_{0}^{k} &= -t(a_{1}^{k} + a_{-1}^{k}) - \tilde{t}_{0}
    b^{k}_{1} \, ,                                  \\
    (\omega - \tilde{\varepsilon}_{d1}) b_{1}^{k} &= - \tilde{t}_{c}
    b_{2}^{k} - \tilde{t}_{0} a^{k}_{0} \, ,         \\
    (\omega - \tilde{\varepsilon}_{d2})b_{2}^{k} &= -\tilde{t}_{c}
    b_{1}^{k} \, .
  \end{align}
\end{subequations}
We can express the amplitudes $b^{k}_{j}$ in terms of $a_{0}^{k}$,
therefore the equation for $a_{j}^{k}$ can be cast into the form
\begin{equation}
\label{e:newdif2}
  \left( \omega - \delta_{j,0}\,\tilde{\varepsilon} \right)a_{j}^{k}
  = -t(a_{j+1}^{k} + a_{j-1}^{k}) \, ,
\end{equation}
where let us define the effective energy $\tilde{\varepsilon}$ as
\begin{equation}
  \tilde{\varepsilon} \equiv \frac{\tilde{t}_{0}^{2} (\omega
  - \tilde{\varepsilon}_{d2})}{(\omega - \tilde{\varepsilon}_{d1})
  (\omega - \tilde{\varepsilon}_{d2}) - \tilde{t}_{c}^{2}} \, .
\end{equation}
This contains all the information about  the side-attached
quantum-dot molecule. Thus, the problem reduces to one of an
impurity of effective energies $\tilde{\varepsilon}$.

In order to study the solutions of \eqref{e:newdif2}, we assume
that the electrons are described by a plane wave incident from the
far one side with unity amplitude and a reflection amplitude $r$,
and at the far other side by a transmission amplitude $\tau$. That
is,
\begin{subequations}
\begin{align}
  a_{j}^{k} &= \text{e}^{\text{i}k\, j} + r\text{e}^{-\text{i}k
  \, j} \, , \, \left( j<0, k>0 \right), \left( j>0, k<0 \right)
  \, ,   \\
  a_{j}^{k} &= \tau \text{e}^{\text{i}k\, j} \, , \qquad \left(
  j>0, k>0 \right), \left( j<0, k<0 \right)
  \label{solut}
\end{align}
\end{subequations}
Inserting \eqref{solut} into \eqref{e:newdif2}, we get a
inhomogeneous system of linear equations for $\tau$, $r$ and
$a_{j}^{k}$, leading to the following expression
\begin{equation}
\label{t-amplitude}
  \tau =  \frac{(\omega - \tilde{\varepsilon}_{-})(\omega
  - \tilde{\varepsilon}_{+})}{(\omega - \tilde{\varepsilon}_{-})
  (\omega - \tilde{\varepsilon}_{+}) + \text{i} (\omega
  - \tilde{\varepsilon}_{d2}) \tilde{\Gamma}} \, ,
\end{equation}
where the bonding energy ($\tilde{\varepsilon}_{-}$) and
antibonding energy ($\tilde{\varepsilon}_{+}$) are defined by
$\tilde{\varepsilon}_{\pm} = (\tilde{\varepsilon}_{d1}
+\tilde{\varepsilon}_{d2} )/2 \pm \sqrt{(\tilde{\varepsilon}_{d1}
+ \tilde{\varepsilon}_{d2} )/2)^2 + \tilde{t}_{c}^{2}}$ and
$\tilde{\Gamma} = \pi \tilde{t}_{0}^{2} \rho(\omega)$ is the
renormalized coupling between the double quantum-dot and the
quantum wire.
The transmission probability is given by $T = |\tau|^{2}$,
\begin{equation}
  T(\omega) = \frac{[(\omega - \tilde{\varepsilon}_{-})(\omega
  - \tilde{\varepsilon}_{+})]^{2}}{[(\omega - \tilde{
  \varepsilon}_{-})(\omega - \tilde{\varepsilon}_{+})]^{2}
  + [(\omega - \tilde{\varepsilon}_{d2}) \tilde{\Gamma}]^{2}} \, .
\end{equation}
In the limit of zero bias and at zero temperature we obtain the
expression for the linear conductance,
\begin{equation}
\label{e:conduct}
  G = \frac{2e^{2}}{h} T(0) = \frac{2e^{2}}{h} \frac{(
  \tilde{\varepsilon}_{-}\tilde{\varepsilon}_{+})^{2}}{(\tilde{\varepsilon}_{-}
  \tilde{\varepsilon}_{+})^{2} + (\tilde{\varepsilon}_{d2}\tilde{\Gamma}
  )^{2}} \, .
\end{equation}
Additionally, from the amplitudes $b_{1}^k$ and $b_{2}^k$ we can
obtain the local density of states (LDOS) in the quantum dots,
\begin{align}
  \rho_{1} & = \frac{1}{\pi}\frac{\tilde{\Gamma}(\omega
  - \tilde{\varepsilon}_{d2})^{2}}{[(\omega - \tilde{\varepsilon}_{-}
  )(\omega - \tilde{\varepsilon}_{+})]^{2} + [(\omega
  - \tilde{\varepsilon}_{d2}) \tilde{\Gamma}]^{2}} \, ,   \\
  \rho_{2} & = \frac{1}{\pi}\frac{\tilde{\Gamma}
  \tilde{t}_{c}^{2}}{[(\omega - \tilde{\varepsilon}_{-})
  (\omega - \tilde{\varepsilon}_{+})]^{2} + [(\omega
  - \tilde{\varepsilon}_{d2})\tilde{\Gamma}]^{2}} \, .
\end{align}
And then we can calculate the density of states of the DQD
molecule.
\begin{equation}
\label{e:dosDQD}
  \begin{split}
    \rho &= \frac{1}{\pi}\frac{\tilde{\Gamma}(\omega
    - \tilde{\varepsilon}_{d2})^{2}+\tilde{t}_{c}^{2}}{[(\omega -
    \tilde{\varepsilon}_{-})(\omega - \tilde{\varepsilon}_{+})]^{2}
    + [(\omega - \tilde{\varepsilon}_{d2}) \tilde{\Gamma}]^{2}} \, .
  \end{split}
\end{equation}

%
%

\section{Results}

We study a model  with $t = 25 \,\Gamma$,  $t_{0} = 7.07 \,
\Gamma$. From here on we use $\Gamma = \pi t_{0}^{2} \rho_{QW}
(0)$ as the energy unit.

First we consider the situation when the energies are varied
simultaneously by a gate voltage $V_{g}$, i.e. $\varepsilon_{d1} =
\varepsilon_{d2} = V_{g}$.

\begin{figure}[h]
\centering
  \includegraphics[angle=-90, scale=0.3]{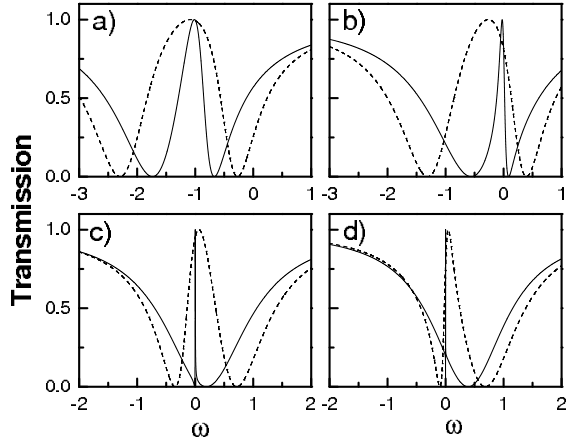}
  \caption{Transmission spectrum for $V_{1g} = V_{2g} = V_{g}=
  -3 \,\Gamma$, $t_{c} = 0.5 \,\Gamma$ (solid line) and $t_{c}
  = \Gamma$ (dashed line), for
  various values of the on site energy $U$, a) $U = 2 \,\Gamma$,
  b) $U = 4 \,\Gamma$, c) $U = 8 \,\Gamma$ and d) $U = 16 \,
  \Gamma$.}
  \label{f:dfkfig2}
\end{figure}

Figure~\ref{f:dfkfig2} displays the transmission probability $T$
for various values of  $U$. For all values of $U$ the transmission
probability reaches the minimum values 0 at $\omega =
\tilde{\varepsilon}_{-}$ and $\tilde{\varepsilon}_{+}$ and its
maximum value the unity at $\omega = \tilde{\varepsilon}_{d2}$.
With increasing $U$, a sharp feature develops close to the Fermi
energy ($\omega = 0$), which is sign of a Kondo resonance. In
fact, for $U$ sufficiently large the transmission can be written
approximately as the superposition of two Fano-Kondo line shapes,
one with a zero $q$ factor and another one with a $q$ factor no
null,
\begin{equation}
  T(\omega) \approx \frac{(\epsilon'+ q')^{2}}{(\epsilon')^{2}
  + 1} + \frac{(\epsilon'' + q'')^2}{(\epsilon'')^{2} + 1}
  \, ,
\end{equation}
where $\epsilon' = (\omega - \tilde{\varepsilon}_{d2})/
\tilde{\Delta}$, $q' = (\tilde{\varepsilon}_{d2} -
\tilde{\varepsilon}_{-})/\tilde{\Delta}$, $\epsilon'' = (\omega -
\tilde{\varepsilon}_{+})/\Gamma$ and $q'' = 0$, with
$\tilde{\Delta} = \tilde{t}_{c}^{2}/\tilde{\Gamma}$.

\begin{figure}[h]
\centering
  \includegraphics[angle=-90, scale=0.3]{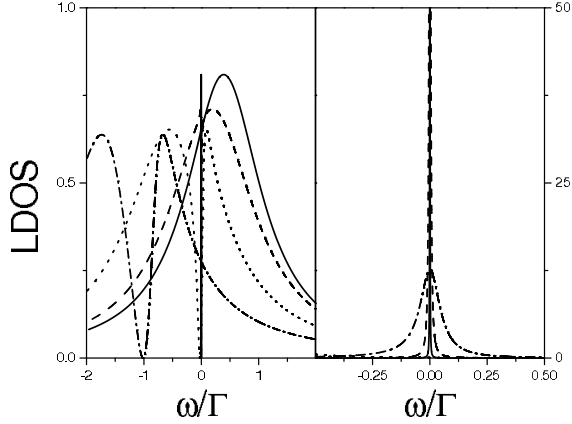}
  \caption{Local density of states a) quantum-dot 1, b)
  quantum-dot $2$, $V_{1g}= V_{2g}=-3\,\Gamma$, for various
  values of the on site energy, $U = 16\,\Gamma$ (solid line),
  $U = 8\,\Gamma$ (dashed line), $U = 4\,\Gamma$ (doted line) and
  $U = 2\,\Gamma$ (dash-dot line).}
  \label{f:dfkfig4}
\end{figure}

The density of states of the quantum-dot molecule can give us a
better understanding of the transport properties of the system.
Figure~\ref{f:dfkfig4} displays DOS for various values of $U$. We
can observe that the density of states is the superposition of a
narrow Kondo peak and a broad Kondo peak with a width that tends
to $2\,\tilde{\Gamma}$. In fact, from Eq.~\eqref{e:dosDQD} the
density of states can be written as the superposition of two
Lorentzian,
\begin{equation}
  \rho \approx \frac{1}{\pi}\frac{\tilde{\Gamma}}{(\omega -
  \tilde{\varepsilon}_{+})^{2} + \tilde{\Gamma}^2} +
  \frac{1}{\pi} \frac{\tilde{\Delta}}{(\omega - \tilde{
  \varepsilon_{d2}})^{2} + \tilde{\Delta}^{2}} \, .
\end{equation}


\begin{figure}[h]
\centering
  \vspace{20pt}
  \includegraphics[angle=-90, scale=0.3]{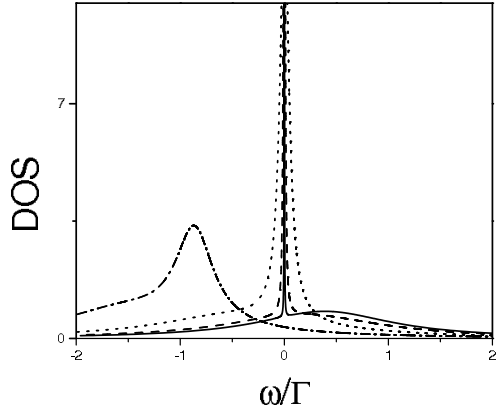}
  \caption{DOS for $t_{c} = 0.5 \,\Gamma$, $V_{1g} = V_{2g} =
  -3 \,\Gamma$, for various values of the on site energy,
  $U = 16 \,\Gamma$ (solid line), $U = 8\,\Gamma$ (dashed line),
  $U = 4\,\Gamma$ (doted line) and $U = 2\,\Gamma$ (dash-dot line).}
  \label{f:dfkfig3}
\end{figure}

Additionally, the LDOS give us more details about the formation of
the Kondo resonance. The LDOS are displayed in
Figure~\ref{f:dfkfig4}. The local density of states $\rho_{2}$
develops a Kondo resonance peak near the Fermi energy ($\omega =
0$). As the outside dot (dot 2) develops a Kondo resonance, the
density of states $\rho_{1}$ is depleted near to the Fermi energy.
From the LDOS we can see that the Kondo resonance is developed in
the outside dot. This behavior is explained by the fact that the
outside dot is weakly connected to the continuum and therefore it
is favorable the formation of a localized state there. The
coherent coupling of this localized state and the conduction
states give origin to this Kondo resonance.

\begin{figure}[h]
\centering
  \includegraphics[angle=-90, scale=0.3]{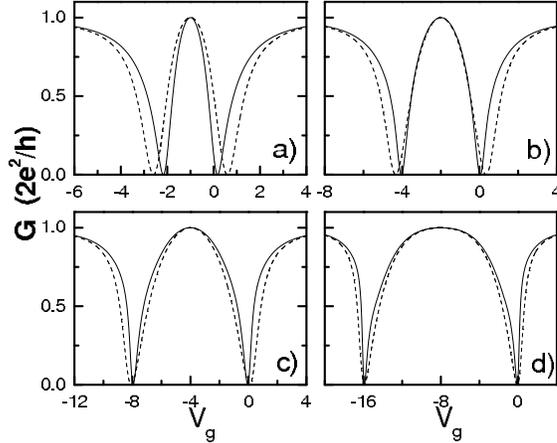}
  \caption{Linear conductance $G$ vs $V_{g}$ for $t_{c} =
  0.5\,\Gamma$ (solid line) and $t_{c} = \Gamma$, for various
  values of the on site energy, a) $U = 2\,\Gamma$, b) $U = 4
  \,\Gamma$, c) $U = 8\,\Gamma$ and d) $U = 16\,\Gamma$.}
  \label{f:dfkfig5}
\end{figure}

Figure~\ref{f:dfkfig5} display the linear conductance versus the
gate voltage, when the two gates voltages are varied
simultaneously ($\varepsilon_{d1} = \varepsilon_{d2} = V_{g}$),
for various values of the on-site energy $U$. The linear
conductance shows two Fano antiresonances corresponding to the
bonding and antibonding energies of the quantum-dot molecule and
one resonance between them. The separation between the two
antiresonance grows linearly with $U$. From the equation
\eqref{e:conduct} the conductance vanishes when
$\tilde{\varepsilon}_{-}$ or $\tilde{\varepsilon}_{+}$ coincide
with the Fermi energy ($\omega = 0$). On the other hand the
conductance reaches the unitary limit when
$\tilde{\varepsilon}_{d2}$ crosses the Fermi energy.

\begin{figure}[h]
\centering
  \includegraphics[angle=-90, scale=0.3]{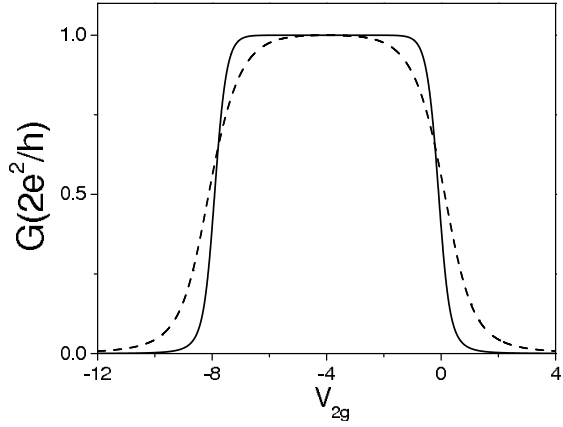}
  \caption{Linear conductance $G$ vs $V_{g}$ for $U = 8\,\Gamma$,
  $t_{c} = 0.5\,\Gamma$ (solid line) and $t_{c} = \Gamma$ (dashed
  line).}
  \label{f:dfkfig6}
\end{figure}

\begin{figure}[h]
\centering
  \includegraphics[angle=-90, scale=0.3]{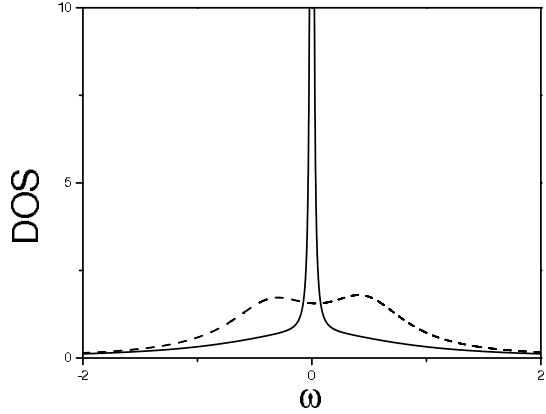}
  \caption{Density of states for $U = 8\,\Gamma$, $V_{1g} =
  -U/2$, $V_{2g} = -U/4$, $t_{c} = 0.5\,\Gamma$ (solid line) and
  $t_{c} = \Gamma$ (dashed line).}
  \label{f:dfkfig7}
\end{figure}

Let us consider now the situation where only one of the gate
potential is varied while the other one is maintained constant
($\varepsilon_{d1} = V_{1g} = -U/2$, $\varepsilon_{d2} = V_{2g}$).
The Figure~\ref{f:dfkfig6} shows the $G$ versus $V_{2g}$ for $U =8
\,\Gamma$. In this situation the occupation number of the quantum
dot 1 (QD1) is $n_{1} \approx 1$ and its energy level is near the
Fermi energy. A Kondo singlet state is formed between the QD1 and
the adjacent wire. For large positive values of $V_{2g}$, the
occupation number of the quantum-dot 2 (QD2) is almost zero and
the position of the antibonding energy, $\tilde{\varepsilon}_{+}
\approx V_{2g}$ and the position of the bonding energy is close to
zero and negative, $\tilde{\varepsilon}_{-} \approx -
\tilde{t}_{c}^{2}/V_{2g} \approx 0$. Consequently, according to
Eq.~\eqref{e:conduct} the conductance is close to zero. Decreasing
$V_{2g}$ the occupation number of the QD2 grows up to $n_{2} = 1$,
and its energy level tends to the Fermi energy. Then the QD2
enters into the Kondo regime and the conductance reaches the
quantum limit ($G = 2e^{2}/h$). The energy level of the QD2 is
pinned at the Fermi energy and the conductance shows a plateau in
its maximum. For $V_{2g} < U$ the QD2 is doubly occupied and a
localized spin-singlet is formed, consequently the Kondo effect of
the coupling of this quantum dot with the quantum wire is
destroyed. On the other hand, QD1 charge is kept almost constant
with one electron and the Kondo singlet state between this quantum
dot and the wire remains. In this situation
$\tilde{varepsilon}_{+}$ tends asymptotically to the Fermi energy
and the conductance decreases to zero.

Figure~\ref{f:dfkfig7} displays the density of states of the
DQD for $V_{2g} = -U/4$,  $t_{c} =0.5\,\Gamma$ (solid line) and
$t_{c} = \Gamma$ (dashed line).  For $t_{c} = 0.5 \,\Gamma$ the
density of states  is the superposition of a broad and a narrow
peaks centered at the Fermi energy with widths $\tilde{\Gamma}$
and $\Delta$ respectively. Then, in this case the side-coupled
DQD has two Kondo temperatures, $T_{1K} = \tilde{\Gamma}$ and
$T_{2K}= \Delta = \tilde{t}_{c}^{2}/\tilde{\Gamma}$. On the other
hand, for $t_{c} = \Gamma$ the density of states shows two peaks
at the bonding and antibonding energies. Both QDs tend to form a
coherent spin-singlet state. This state coexists with the Kondo
states formed between the QD and the QW.

%
%
\section{Summary}

In conclusion, we have studied the transport through a
side-coupled double quantum-dot molecule using the finite-$U$
slave boson mean field approach. We have found that the
transmission spectrum shows a structure with two antiresonances
localized at the bonding and antibonding energies of the
quantum-dot molecule, and one resonance at the site energy of the
outside quantum-dot. Moreover the density of states shows that the
outside quantum-dot develops a strong Kondo effect with the
quantum-wire and its Kondo temperature depends strongly on the
interdot coupling tunneling. The linear conductance reflects the
transmission spectrum properties as the gate potential is varied.

%
%
\section*{Acknowledgments}
  The authors would like to thank for financial support: G.A.L. and
P.A.O. to thank Milenio ICM P02-054F, P.A.O. to thank FONDECYT
(grants 1020269 and 7020269), and J.M.Y. and G.A.L. to thank U.A.
(PEI-1305-04).

%

%
%

\begin{thebibliography}{10}
\expandafter\ifx\csname url\endcsname\relax
\def\url#1{\texttt{#1}}\fi \expandafter\ifx\csname
urlprefix\endcsname\relax\def\urlprefix{URL }\fi

\bibitem{JHW98}
L.~Jacak, P.~Hawrylak, A.~W{\'o}js, Quantum {D}ots,
Springer--Verlag, Berlin, 1998.

\bibitem{K69}
J.~Kondo, Theory of dilute magnetic alloys, in: F.~Seitz,
D.~Turnbull, H.~Ehrenreich (Eds.), Solid {S}tate {P}hysics,
Vol.~23, Academic Press, New York, 1969, p. 184.

\bibitem{K64}
J.~Kondo, Resistance minimum in dilute magnetic alloys, Prog.
Theor. Phys. 32~(1) (1964) 37--69.

\bibitem{H97}
A.~C. Hewson, The {K}ondo {P}roblem to {H}eavy {F}ermions, Vol.~2
of Cambridge studies in magnetism, Cambridge University Press,
Cambridge, 1997.

\bibitem{GSMAMK98}
D.~Goldhaber-Gordon, H.~Shtrikman, D.~Mahalu, D.~Abusch-Magder,
U.~Meirav, M.~A. Kastner, Kondo effect in a single-electron
transistor, Nature 391~(6663) (1998) 156--159.

\bibitem{GGKSMM98}
D.~Goldhaber-Gordon, J.~G{\"o}res, M.~A. Kastner, H.~Shtrikman,
D.~Mahalu, U.~Meirav, From the {K}ondo regime to the mixed-valence
regime in a single-electron transistor, Phys. Rev. Lett. 81~(23)
(1998) 5225–--5228.

\bibitem{COK98}
S.~M. Cronenwett, T.~H. Oosterkamp, L.~P. Kouwenhoven, A tunable
{K}ondo effect in quantum dots, Science 281~(5376) (1998)
540--544.

\bibitem{GR88}
L.~I. Glazman, M.~{\'E}. Raikh, Resonant {K}ondo transparency of a
barrier with quasilocal impurity states, JETP Lett. 47~(8) (1988)
452--455.

\bibitem{NL88}
T.~K. Ng, P.~A. Lee, On-site {C}oulomb repulsion and resonant
tunneling, Phys. Rev. Lett. 61~(15) (1988) 1768–--1771.

\bibitem{PG04}
M.~Pustilnik, L.~Glazman, Kondo effect in quantum dots, J. Phys.:
Condens. Matter 16~(16) (2004) R513--R537.

\bibitem{GM99}
A.~Georges, Y.~Meir, Electronic correlations in transport through
coupled quantum dots, Phys. Rev. Lett. 82~(17) (1999) 3508--–3511.

\bibitem{AE01}
T.~Aono, M.~Eto, Kondo resonant spectra in coupled quantum dots,
Phys. Rev. B 63~(12) (2001) 125327.

\bibitem{JCM01}
H.~Jeong, A.~M. Chang, M.~R. Melloch, The {K}ondo effect in an
artificial quantum dot molecule, Science 293~(5538) (2001)
2221--2223.

\bibitem{SG02}
Q.~feng Sun, H.~Guo, Double quantum dots: {K}ondo resonance
induced by an interdot interaction, Phys. Rev. B 66~(15) (2002)
155308.

\bibitem{LOA03}
G.~A. Lara, P.~A. Orellana, E.~V. Anda, Tristability in a
non-equilibrium double-quantum-dot in the {K}ondo regime, Solid
State Comm. 125~(3--4) (2003) 165--169.

\bibitem{DRVRPM00}
P.~Debray, O.~E. Raichev, P.~Vasilopoulos, M.~Rahman, R.~Perrin,
W.~C. Mitchell, Ballistic electron transport in stubbed quantum
waveguides: Experiment and theory, Phys. Rev. B 61~(16) (2000)
10950--10958.

\bibitem{TIK02}
Y.~Takazawa, Y.~Imai, N.~Kawakami, Electron {T}ransport through
{T}-{S}haped {D}ouble-{D}ots {S}ystem, J. Phys. Soc. Jpn. 71~(9)
(2002) 2234--2239.

\bibitem{GSG03}
A.~D. G{\"u}{\c{o}}l{\"u}, Q.~F. Sun, H.~Guo, Kondo resonance in a
quantum dot molecule, Phys. Rev. B 68~(24) (2003) 245323.

\bibitem{KCKS01}
K.~Kang, S.~Y. Cho, J.-J. Kim, S.-C. Shin, Anti-{K}ondo resonance
in transport through a quantum wire with a side-coupled quantum
dot, Phys. Rev. B 63~(11) (2001) 113304.

\bibitem{THCP02}
M.~E. Torio, K.~Hallberg, A.~H. Ceccatto, C.~R. Proetto, Kondo
resonances and {F}ano antiresonances in transport through quantum
dots, Phys. Rev. B 65~(8) (2002) 085302.

\bibitem{ODGL03}
P.~A. Orellana, F.~Dom\'{\i}nguez-Adame, I.~G\'{o}mez, M.~L.~L.
de~Guevara, Transport through a quantum wire with a side
quantum-dot array, Phys. Rev. B 67~(8) (2003) 085321.

\bibitem{FFA03}
R.~Franco, M.~S. Figueira, E.~V. Anda, Fano resonance in
electronic transport through a quantum wire with a side-coupled
quantum dot: {X}-boson treatment, Phys. Rev. B 67~(15) (2003)
155301.

\bibitem{S03}
P.~Stefa{\'n}ski, Quantum dots as scatterers in electronic
transport: interference and correlations, Solid State Comm.
128~(1) (2003) 29--34.

\bibitem{GGHK00}
J.~G{\"o}res, D.~Goldhaber-Gordon, S.~Heemeyer, M.~A. Kastner,
Fano resonances in electronic transport through a single-electron
transistor, Phys. Rev. B 62~(3) (2000) 2188--2194.

\bibitem{ZGGKKSMM01}
I.~G. Zacharia, D.~Goldhaber-Gordon, G.~Granger, M.~A. Kastner,
Y.~B. Khavin, H.~Shtrikman, D.~Mahalu, U.~Meirav, Temperature
dependence of {F}ano line shapes in a weakly coupled
single-electron transistor, Phys. Rev. B 64~(15) (2001) 155311.

\bibitem{F61}
U.~Fano, Effects of {C}onfiguration {I}nteraction on {I}ntensities
and {P}hase {S}hifts, Phys. Rev. 124~(6) (1961) 1866--1878.

\bibitem{TB93}
E.~Tekman, P.~F. Bagwell, Fano resonances in quasi-one-dimensional
electron waveguides, Phys. Rev. B 48~(4) (1993) 2553–--2559.

\bibitem{NS94}
J.~U. N{\"{o}}ckel, A.~D. Stone, Resonance line shapes in
quasi-one-dimensional scattering, Phys. Rev. B 50~(23) (1994)
17415–--17432.

\bibitem{KR86}
G.~Kotliar, A.~E. Ruckenstein, New functional integral approach to
strongly correlated fermi systems: The {G}utzwiller approximation
as a saddle point, Phys. Rev. Lett. 57~(11) (1986) 1362--1365, and
references cited therein.

\bibitem{DL01}
B.~Dong, X.~L. Lei, Kondo-type transport through a quantum dot
under magnetic fields, Phys. Rev. B 63~(23) (2001) 235306.

\bibitem{DL01b}
B.~Dong, X.~L. Lei, Kondo-type transport through a quantum dot: a
new finite-${U}$ slave-boson mean-field approach, J. Phys.:
Condens. Matter 13~(41) (2001) 9245--9258.

\bibitem{DL02}
B.~Dong, X.~L. Lei, Kondo effect and antiferromagnetic correlation
in transport through tunneling-coupled double quantum dots, Phys.
Rev. B 65~(24) (2002) 241304(R).

\bibitem{DL02b}
B.~Dong, X.~L. Lei, Nonequilibrium {K}ondo effect in a multilevel
quantum dot near the singlet-triplet transition, Phys. Rev. B
66~(11) (2002) 113310.

\bibitem{B76}
S.~E. Barnes, New method for the {A}nderson model, J. Phys. F:
Met. Phys. 6~(7) (1976) 1375--1383.

\bibitem{B77}
S.~E. Barnes, New method for the {A}nderson model. ii. the ${U} =
0$ limit, J. Phys. F: Met. Phys. 7~(12) (1977) 2637--2647.

\bibitem{C84}
P.~Coleman, New approach to the mixed-valence problem, Phys. Rev.
B 29~(6) (1984) 3035--–3044.

\end{thebibliography}
\end{document}